\documentclass[aps,10pt,prd,preprintnumbers,twocolumn,nofootinbib,notitlepage,superscriptaddress,
floatfix]{revtex4-2}

\usepackage{graphicx}
\usepackage{dcolumn}
\usepackage{bm}
\usepackage{amsmath,amssymb}
\usepackage{slashed}
\usepackage{xcolor}
\usepackage{physics}
\usepackage{multirow}
\usepackage{mathtools,braket}
\usepackage{soul}
\usepackage{lipsum}  
\definecolor{blue}{rgb}{0.0, 0.0, 1.0}
\definecolor{red}{rgb}{1.0, 0.0, 0.0}
\definecolor{royalblue}{rgb}{0.0, 0.14, 0.4}

\usepackage{hyperref}
\hypersetup{colorlinks=true,citecolor=blue,linkcolor=blue,urlcolor=blue}

\usepackage[mathlines]{lineno}
\usepackage{mathrsfs}
\usepackage{tikz}
\definecolor{lime}{HTML}{A6CE39}
\DeclareRobustCommand{\orcidicon}{%
	\begin{tikzpicture}
	\draw[lime, fill=lime] (0,0) 
	circle [radius=0.16] 
	node[white] {{\fontfamily{qag}\selectfont \tiny ID}};
	\draw[white, fill=white] (-0.0625,0.095) 
	circle [radius=0.007];
	\end{tikzpicture}
	\hspace{-2mm}
}
\foreach \x in {A, ..., Z}{%
	\expandafter\xdef\csname orcid\x\endcsname{\noexpand\href{https://orcid.org/\csname orcidauthor\x\endcsname}{\noexpand\orcidicon}}
}

\begin{document}
%----------------------------------------------------------------
\preprint{LFTC-26-06/115}
%----------------------------------------------------------------
%================================================================
\title{Electromagnetic structure of strange vector mesons in nuclear medium}
%================================================================

\author{Parada~T.~P.~Hutauruk\orcidA{}} 
\email[E-mail:]{phutauruk@hiroshima-u.ac.jp}
\affiliation{International Institute for Sustainability with Knotted Chiral Meta Matter (WPI-SKCM$^2$), Hiroshima University, Higashi-Hiroshima, Hiroshima 739-8526, Japan}

\author{Terry Mart\orcidC{}}
\email[E-mail: ]{terry.mart@sci.ui.ac.id}
\affiliation{Departemen Fisika, FMIPA, Universitas Indonesia, Depok 16424, Indonesia}

\author{Kazuo Tsushima\orcidB{}}
\email[E-mail: ]{kazuo.tsushima@gmail.com} 
\affiliation{Laboratório de Física Teórica e Computacional-LFTC, Programa de P\'{o}sgradua\c{c}\~{a}o em Astrof\'{i}sica e F\'{i}sica Computacional,
Universidade Cidade de S\~{a}o Paulo, 01506-000 S\~{a}o Paulo, SP, Brazil}\textbf{}

\date{\today}

%================================================================
\begin{abstract} 
We investigate the in-medium modifications of the charge (electric) $G_C^{*}(Q^2)$,
magnetic $G_M^{*}(Q^2)$, and quadrupole $G_Q^{*}(Q^2)$ form factors of the positively charged vector meson $K^{*+}$ in symmetric nuclear matter at zero temperature within
the Schwinger proper-time Nambu-Jona-Lasinio (NJL) model. In this framework, both the
nuclear medium effects and the electromagnetic structure of the $K^{*+}$ meson are
described consistently in the NJL model at the quark level. We find that the charge, magnetic, and
quadrupole form factors are suppressed with increasing nuclear density, indicating substantial
in-medium modifications of the strange vector meson's internal structure.
We further obtain a charge radius of $r_{K^{*+}}^{*}=0.74~\mathrm{fm}$ at normal nuclear density,
which is slightly smaller than the corresponding value for the $\rho^{+}$ meson,
$r_{\rho^{+}}^{*}\simeq0.75~\mathrm{fm}$.
\end{abstract}
\maketitle

%================================================================
\section{Introduction} \label{sec:intro}
%================================================================
The electromagnetic form factor (EMFF) is one of the primary observables for probing the internal
structure of hadrons. It encodes information on the spatial distributions of electric
charge and magnetization, thereby providing valuable insight into the underlying quark dynamics.
Among hadronic systems, the strange vector meson $K^{*+}$, with spin $J=1$, may exhibit a
distinctive internal structure compared with both the $\rho^+$ vector meson and the pseudoscalar
mesons $\pi^+$ and $ K^{+} $.
Although the $K^{*+}$ and $\rho^+$ vector mesons share the same spin quantum number, they differ in
their quark flavor composition, making the comparison particularly useful for
investigating flavor SU(3) symmetry breaking. In contrast, the $K^{+}$ pseudoscalar and
$K^{*+}$ vector mesons possess the same valence quark content but differ in their spin quantum
numbers, with the $K^+$ meson being a pseudoscalar with $J=0$ and the $K^{*+}$ a vector meson with
$J=1$. This difference in spin leads to distinct spin and orbital angular momentum correlations,
resulting in markedly different partonic structures despite their identical flavor quark content.
Consequently, the electromagnetic structure of the $K^{*+}$ meson provides a unique opportunity to
investigate the interplay between quark flavor and spin degrees of freedom in hadrons, making it an
important subject in contemporary studies of hadron structure.

From a theoretical perspective, the EMFFs of the $K^{*+}$ and $K^{*0}$ mesons have been
investigated using a variety of theoretical approaches and phenomenological
models~\cite{Brodsky:1992px,Xu:2024fun,Bhagwat:2006pu,Hawes:1998bz,Hutauruk:2026oge,
Miramontes:2025vzb,Lorce:2009bs,DeMelo:2018bim,Carrillo-Serrano:2015uca,Zhang:2024nxl,
Krutov:2018mbu,Haberzettl:2019qpa,Sun:2020jng,Ninomiya:2017ggn,Shi:2023oll,Liu:2025fuf,Choi:2004ww,
Hecht:1997uj,Braguta:2004kx,Aliev:2004uj,Jaus:2002sv,Cloet:2014rja,Roberts:2011wy,
Gutierrez-Guerrero:2026rsb,Luan:2015goa,Luschevskaya:2026kxx,GarciaGudino:2010sd,Mart:1997cc,
Mart:2011ez}.
However, these
investigations have been carried out almost exclusively in free space (vacuum). To date, only a few
studies have examined the in-medium EMFFs of the $\rho^+$ vector
meson~\cite{Hutauruk:2025bjd,Gautam:2026ohc,deMelo:2018hfw}, while corresponding studies of the
$K^{*+}$ strange vector meson remain unavailable.

The motivation for investigating in-medium modifications stems from the discovery of
the EMC effect by the European Muon Collaboration (EMC)~\cite{EuropeanMuon:1983wih}, which
demonstrated that the nucleon structure functions measured in nuclei differ significantly from those
of free nucleons. This observation established that the internal structure of hadrons is modified in
the nuclear medium. Subsequent theoretical and experimental studies have
further shown that hadron properties, including masses, decay constants, and electromagnetic
observables, also undergo in-medium
modifications~\cite{Hayano:2008vn,Hatsuda:1991ez,Gifari:2024ssz,Balassa:2025gwv,Sasaki:2022vas,
Mutuk:2025lak}. Consequently, the investigation of in-medium hadron structure has become an
active area of research in theory and experiment~\cite{Hen:2016kwk,Paakkinen:2026dnh,Deur:2025inv}.
Nevertheless, most previous studies have focused on the nucleon and light vector mesons. In
contrast, the in-medium electromagnetic structure of the $K^{*+}$ meson remains
largely unexplored. Therefore, in this work, we investigate the in-medium modifications
of the electromagnetic structure of the $K^{*+}$ meson.

On the experimental side, the BABAR Collaboration has measured the cross section for
the reaction $e^+e^- \rightarrow K^+K^-\pi^0\pi^0$~\cite{BaBar:2011btv}. A recent
analysis~\cite{Rojas:2024tmn} demonstrated that this cross section is sensitive to the magnetic
moment of the $\rho^+$ vector meson, $\mu_{\rho^+}(Q^2)$, where $Q^2$ denotes the virtual-photon
momentum transfer~\cite{Rojas:2024tmn,Dbeyssi:2011ep}. A similar analysis has also been performed
for the $K^{*+}$ meson~\cite{Perez:2026yxg}. In addition, the PANDA experiment will
investigate antiproton-induced reactions that provide access to the electromagnetic structure of
vector mesons, including processes such as $\bar{p}p \rightarrow \rho^-\rho^+$~\cite{PANDA:2009yku}.
Together with measurements from $e^+e^-$ annihilation experiments, these studies will enable the
extraction of the timelike EMFFs of vector mesons and provide valuable constraints on theoretical
models. Future high-precision measurements from experiments such as BABAR and PANDA will therefore
offer stringent tests of theoretical predictions for the EMFFs of vector mesons over a broad range
of momentum transfer.

Beyond the measurements performed by the BABAR Collaboration, future experimental programs
will further improve our understanding of the vector-meson electromagnetic structure. The Belle II
experiment is expected to provide high-precision measurements of exclusive $e^+ e^-$ annihilation
processes involving vector mesons, thereby constraining their timelike
EMFFs~\cite{Belle-II:2025wpi}. Complementary information on the spacelike electromagnetic structure
of vector mesons can be obtained from exclusive electroproduction experiments at Jefferson
Lab~\cite{CLAS:2001zwd,Strikman:2008pi,Mart:2026tyg}, while the future Electron-Ion Collider (EIC)~\cite{Boer:2011fh,Accardi:2012qut,Aschenauer:2017jsk} will enable precision
studies of exclusive vector-meson production from both nucleon and nuclear targets, providing unique
insight into the partonic structure of vector mesons and their in-medium modifications. This broad
experimental program will offer stringent tests of theoretical predictions for vector-meson EMFFs.

In this paper, we investigate the in-medium modifications of the EMFFs of the positively charged meson $K^{*+}$, namely, the charge (electric) $G_C^{*}(Q^2)$, magnetic
$G_M^{*}(Q^2)$, and quadrupole $G_Q^{*}(Q^2)$ form factors, together with the corresponding charge
radius. This work extends our previous studies of the vacuum electromagnetic properties of the
$K^{*+}$ meson presented in Refs.~\cite{Hutauruk:2025bjd,Hutauruk:2026oge} to the
case of symmetric nuclear matter (SNM). The NJL model has been successfully applied to a wide range of
hadronic and nuclear phenomena, including the gluon distribution in nuclear
matter~\cite{Hutauruk:2021kej}, the phase structure of quark matter and neutron
matter~\cite{Whittenbury:2015ziz}, the properties of neutron stars~\cite{Bentz:2001vc}, the EMFFs of
diquarks and nucleons~\cite{Cloet:2014rja}, and charge symmetry breaking in EMFFs and parton
distribution functions~\cite{Hutauruk:2018zfk}. Motivated by these successful applications, we
employ the NJL model to investigate the in-medium electromagnetic structure of the $K^{*+}$ meson. Finally, we compare our results with %available lattice QCD simulations and 
other theoretical calculations.

This paper is organized as follows. In Sec.~\ref{sec:vacuumNJL}, we briefly review
the description of the $K^{*+}$ meson within the covariant NJL model employing the
Schwinger proper-time regularization scheme to simulate quark confinement. In Sec.~\ref{sec:ffv}, we
formulate the in-medium EMFFs of the $K^{*+}$ meson from the corresponding
electromagnetic current matrix elements and derive the medium-modified charge (electric)
$G_C^{*}(Q^2,\rho/\rho_0)$, magnetic $G_M^{*}(Q^2,\rho/\rho_0)$, and quadrupole
$G_Q^{*}(Q^2,\rho/\rho_0)$ form factors with $\rho$ $(\rho_0)$ being nuclear (normal nuclear) density,
together with the corresponding charge radius,
$r_{K^{*+}}^{*}$. In Sec.~\ref{sec:MR}, we present and discuss our numerical results for the
in-medium EMFFs of the $K^{*+}$ meson and the associated static electromagnetic
properties. Finally, our summary and conclusion are presented in Sec.~\ref{sec:summary}.

%================================================================
\section{Three-flavor NJL model} 
\label{sec:vacuumNJL}
%================================================================
In this section, we briefly review the three-flavor NJL model, a Poincaré-covariant effective
quantum field theory that captures the essential low-energy features of quantum chromodynamics
(QCD), most notably dynamical chiral symmetry breaking (DCSB). In the NJL model, DCSB is realized
through the self-consistent solution of the gap equation, which generates nonvanishing chiral
condensates and dynamically dressed quark masses, thereby transforming the current quark
masses into dynamical quark masses. Although the NJL model does not explicitly exhibit
a QCD confinement mechanism, confinement can be effectively simulated by employing the Schwinger
proper-time regularization scheme. This regularization removes the unphysical quark-production
thresholds associated with
hadronic decays into free quarks while preserving the underlying symmetries of the model. The
resulting three-flavor NJL effective Lagrangian, formulated in terms of local four-fermion contact
interactions, is given by
\begin{eqnarray}
    \label{eq:vacNJL1}
    \mathscr{L}_{\mathrm{NJL}} &=& \bar{\psi}_q \big( i \partial\!\!\!/ -\hat{m}_q \big) \psi_q
\nonumber \\
    &+& G_\pi \big[ \big(\bar{\psi}_q \lambda_i \psi_q \big)^2 -\big( \bar{\psi}_q \gamma_5
\lambda_i
    \psi_q \big)^2 \big] \nonumber \\
    &-& G_\rho \big[ \big( \bar{\psi}_q \gamma^\mu \lambda_i \psi_q \big)^2
    + \big( \bar{\psi}_q \gamma^\mu \gamma_5 \lambda_i \psi_q \big)^2 \big],
\end{eqnarray}
where $\psi^{T}=(\psi_u,\psi_d,\psi_s)$ denotes the quark field in flavor space, $\lambda_i$
($i=1,\ldots,8$) are the Gell-Mann matrices with $\lambda_0=\sqrt{2/3}\,\mathbf{1}$, and
$\hat{m}_q=\mathrm{diag}(m_u,m_d,m_s)$ is the current-quark mass matrix. The four-fermion
coupling constant characterizes interaction in the scalar and pseudoscalar channels
$G_\pi$, which drives quark--antiquark correlations and is responsible for DCSB. The coupling
constant $G_\rho$ governs the quark--antiquark interaction in the vector and axial-vector channels.

In the Hartree mean-field approximation, the dressed quark masses are obtained self-consistently
from the quark self-energy by solving the gap equation. Within the Schwinger proper-time
regularization scheme, the gap equation takes the form
\begin{eqnarray}
    \label{eq:vacuumNJL2}
    M_q &=& m_q + \frac{N_c G_\pi M_q}{\pi^2} \int_{\tau_{\mathrm{UV}}}^{\tau_{\mathrm{IR}}}
    \frac{d\tau}{\tau^2} e^{-\tau M_q^2},
\end{eqnarray}
where $\tau_{\mathrm{UV}}=\Lambda_{\mathrm{UV}}^{-2}$ and
$\tau_{\mathrm{IR}}=\Lambda_{\mathrm{IR}}^{-2}$ denote the ultraviolet (UV) and infrared (IR)
proper-time cutoffs, respectively. Throughout this work, the infrared cutoff is fixed at
$\Lambda_{\mathrm{IR}}=0.24~\mathrm{GeV}$, a value chosen to simulate quark confinement, while the
ultraviolet cutoff $\Lambda_{\mathrm{UV}}$ is determined by reproducing the empirical pion mass,
$m_\pi=140~\mathrm{MeV}$, and pion decay constant, $f_\pi=93~\mathrm{MeV}$. The parameters
$\Lambda_{\mathrm{UV}}$ and $\Lambda_{\mathrm{IR}}$ define the regularization scales of the
proper-time scheme and are treated as intrinsic parameters of the NJL model. The dressed quark
propagator for a quark of flavor $q=(\ell,s)$ is then given by
\begin{eqnarray}
    \label{eq:gap1}
    S_{\ell} (k) &=& \frac{(k\!\!\!/ - M_{\ell})}{(k^2 - M_{\ell}^2 + i \epsilon)},  \\
    \label{eq:gap1a}
    S_s (k) &=& \frac{(k\!\!\!/ - M_s)}{(k^2 - M_s^2 + i \epsilon)},
\end{eqnarray}
where the subscripts $\ell=(u,d)$ and $s$ refer to the light- and strange-quark sectors, respectively.

The dressed quark--antiquark bound state corresponding to the $K^{*+}$ meson
is obtained by solving the homogeneous Bethe--Salpeter equation (BSE) in the random-phase
approximation. In the vector-meson channel, the solution is equivalently described by the
quark--antiquark scattering $t$-matrix. Summing the quark--antiquark bubble diagrams to all orders
generates a geometric series, yielding the reduced form of the following $t$-matrix, and it gives
\begin{eqnarray}
    \label{eq:vacuumNJL3}
    t_{K^{*+}}^{\mu \nu} (p^2) &=&  \frac{-4iG_{\rho}}{1 + 2 G_{\rho} \Pi_{K^{*+}} (p^2)} 
    \nonumber \\
    &\times& \Bigg( g^{\mu \nu} + 2 G_{\rho} \Pi_{K^{*+}} (p^{2}) \frac{p^{\mu} p^{\nu}}{p^2}\Bigg).
\end{eqnarray}
The polarization insertions (bubble diagrams) for the $K^{*+}$ meson are given by
\begin{eqnarray}
    \label{eq:vacuumNJL4}
%    \MoveEqLeft 
      &&\Pi_{K^{*+}} (p^2) P^{\mu \nu} \delta_{ab} \nonumber\\
      && = iN_c \int \frac{d^4k}{(2\pi)^4} %\nonumber \\ &\times& 
      \mathrm{Tr} \big[ \gamma^\mu \lambda_a S_{\ell} (p+k) \gamma^\nu \lambda_b S_{s} (k)
      \big],~~
\end{eqnarray}
where $P^{\mu\nu}= (g^{\mu\nu}-p^\mu p^\nu/p^2$) denotes the transverse projection operator,
and $S_{\ell}(p+k)$ and $S_s(k)$ represent the dressed light- and strange-quark propagators,
respectively, as defined in Eqs.~(\ref{eq:gap1}) and~(\ref{eq:gap1a}).
The trace over color degrees of freedom has already been carried out, while the remaining trace is
taken over Dirac and flavor indices. Within the Schwinger proper-time regularization scheme, the
polarization insertion for the
$K^{*+}$ meson can be expressed in the following form
\begin{eqnarray}
    \Pi_{K^{*+}} (p^2) &=& \frac{N_c}{2 \pi^2} \int_0^1 dx
    \int_{\tau_{\mathrm{UV}}}^{\tau_{\mathrm{IR}}} \frac{d\tau}{\tau} \exp \big[ -\tau
(C_1) \big],\nonumber \\
    &\times& \Big[ M_uM_s -(1-x)M_u^2 -xM_s^2 \nonumber \\
    &+&  2x(1-x) p^2 \Big],
\end{eqnarray}
where $C_1 =\left[xM_u^2 + (1-x) M_s^2-x(1-x)p^2\right]$.
The $K^{*+}$ meson mass can be determined straightforwardly from the pole position of
the corresponding $t$-matrix in Eq.~(\ref{eq:vacuumNJL3}), which yields
\begin{eqnarray}
    \label{eq:vacuumNJL5}
    1 + 2 G_\rho \Pi_{K^{*+}} \big( p^2 = m_{K^{*+}}^2\big) &=& 0,
\label{ksmass}
\end{eqnarray}
where the relevant quantities are evaluated at the bound-state pole of the $t$-matrix. In Eq.~(\ref{ksmass}), $m_{K^{*+}}$ denotes the mass of the $K^{*+}$ meson. To determine the $K^{*+}$--quark coupling constant, we expand the $t$-matrix
in Eq.~(\ref{eq:vacuumNJL3}) around the bound-state pole, $p^2=m_{K^{*+}}^2$, corresponding to the
on-shell condition. The meson wave-function renormalization constant, which is directly related to
the meson--quark coupling constant, is then obtained as
\begin{eqnarray}
    \label{eq:vacuumNJL8}
   Z_{K^{*+}}^{-1} = \big[ g_{K^{*+} qq}\big]^{-2} &=& - \frac{\partial \Pi_{K^{*+} qq} (p^2)}{\partial p^2} \Bigg|_{p^2 = m_{K^{*+}}^2},
\end{eqnarray}
where $g_{K^{*+} qq}$ denotes the $K^{*+}$ meson–quark coupling constant. Its expression in the Schwinger proper-time regularization scheme can be written as
\begin{eqnarray}
 \label{eq:vacuumNJL9}
    \big[g_{K^{*+} qq}\big]^{-2} &=& \frac{N_c}{2 \pi^2} \int_{\tau_{\mathrm{UV}}}^{\tau_{\mathrm{IR}}} d\tau \int_0^1  dx x(1-x) \exp \big[-\tau (C_2) \big] \nonumber \\
    &\times& \Big[ M_u M_s -(1-x)Mu^2 -xM_s^2 \nonumber \\
    &-& 2x(1-x) m_{K^{*}}^2 +\frac{2}{\tau} \Big], 
\end{eqnarray}
with $C_2 = \left[xM_u^2 + (1-x) M_s^2 -x(1-x) m_{K^{*+}}^2\right]$.

%================================================================
\section{Nuclear matter in the NJL model} \label{sec:NMNJL}
%================================================================
We extend the NJL model Lagrangian to SNM by applying
the Fierz transformation~\cite{Bentz:2001vc}. Under this transformation, the four-fermion
interaction terms are rearranged into a chiral-symmetric linear combination of the form $ \sum_i
G_i\left(\bar{\psi}_q\Gamma_i\psi_q\right)^2$,
where $\Gamma_i$ denotes the relevant Dirac and flavor structures of the interaction channels
(see Ref.~\cite{Bentz:2001vc} for details). Applying the Fierz transformation to the quark bilinears
in the effective NJL Lagrangian of Eq.~(\ref{eq:vacNJL1}), we
obtain~\cite{Bentz:2001vc,Gifari:2024ssz}
\begin{eqnarray} 
\label{eq:njl1}
\mathscr{L}_{\mathrm{NM-NJL}} &=& \bar{\psi}_q \big( i \partial\!\!\!/ - M_q - V\!\!\!\!/
\big)\psi_q \nonumber \\
&-& \frac{\big( M_q-m_q\big)^2}{4G_\pi} + \frac{V_\mu V^\mu}{2G_\omega} + \mathscr{L}_{I},
\end{eqnarray}  
where $\mathscr{L}_{I}$ denotes the interaction Lagrangian. By applying the Fierz transformation
and charge conjugation, the interaction Lagrangian $\mathscr{L}_{I}$ in Eq.~(\ref{eq:njl1}) can be
rearranged into a sum of the isoscalar-scalar ($T=0, J^P=0^+$) and isovector-axial-vector ($T=1, J^P=1^+$) diquark interaction
channels. Accordingly, the interaction Lagrangian can be expressed as
\begin{eqnarray}
\mathscr{L}_{I,qq} &=& G_s \big[ \bar{\psi}_q \gamma_5 C \tau_2 \beta_A \bar{\psi}_q^T \big] \big[
\psi^T C^{-1} \gamma_5 \tau_2 \beta_A \psi_q \big] \nonumber \\
&+& G_a \big[ \bar{\psi}_q \gamma_\mu C \tau_i
\tau_2 \beta_A \bar{\psi}_q^T \big] \big[ \psi_q^T C^{-1} \gamma^\mu \tau_2 \tau_i \beta_A \psi_q
\big], \nonumber \\
\end{eqnarray}
where $\beta_A=\sqrt{\frac{3}{2}}\,\lambda_A$ with $A=2,5,7$, and $C=i\gamma_2\gamma_0$ is
the charge-conjugation matrix. The coupling constants $G_s$ and $G_a$ represent the strengths of
the scalar and axial-vector diquark interactions, respectively. The scalar diquark coupling
constant $G_s$ is determined by fitting the free nucleon mass. The isoscalar-vector mean field and
the dynamically generated effective quark mass are defined as
\begin{eqnarray}
\label{eq:NJL8}
V^\mu &=& 2 G_\omega \big< \rho_B \mid \bar{\psi}_q \gamma^\mu \psi_q \mid \rho_B \big> \nonumber \\
&=& 2
\delta^{0\mu}
G_\omega \big< \psi_q^\dagger \psi_q \big>, \\
\label{eq:NJL8b}
M_q &=& m_q - 2 G_\pi \big< \rho_B \mid \bar{\psi}_q \psi_q \mid \rho_B \big>,
\label{Mq}
\end{eqnarray}  
where the vector mean field is defined as $V^\mu=( V^0=V_0,\mathbf{0})$ for SNM in its rest frame. By imposing the stability (minimum-energy) condition, $\partial
\mathcal{E}/\partial V_0=0$, the vector mean field is determined as $V_0=6G_\omega\rho_B$,
where $\rho_B=2p_F^3/(3\pi^2)$ denotes the baryon number density of SNM.
For a fixed baryon density, the dynamically generated in-medium quark mass $M_q$ is obtained
by requiring the corresponding thermodynamic potential to satisfy the stationary condition,
$ {\partial\mathscr{E}}/{\partial M_q}
= {(M_q-m_q)}/{2G_\pi} +\langle \rho_B|\bar{\psi}_q\psi_q|\rho_B\rangle =0$,
which determines the in-medium quark mass consistently with Eq.~(\ref{eq:NJL8}).
Employing the standard hadronization procedure~\cite{Bentz:2001vc}, the effective potential for SNM can be derived from the effective NJL Lagrangian. Within the mean-field
approximation, the energy density of SNM is then expressed as
\begin{eqnarray}
\label{eq:NJL9} 
\mathscr{E} = \mathscr{E}_V- \frac{V_0^2}{4G_\omega} + 4 \int \frac{d^3p}{(2\pi)^3} \Theta
\big(p_F - |\mathbf{p}|\big)\, \epsilon_N,
\end{eqnarray}  
where $\epsilon_N=\sqrt{M_N^{*2}+\mathbf{p}^2}+3V_0\equiv E_N+3V_0$ with $M_N^*$ denoting the in-medium nucleon mass obtained from the pole of
the quark--diquark $t$-matrix. The nucleon Fermi momentum is determined by
$
p_F^2=\left(\mu_N-3V_0\right)^2-M_N^{*2},
$
where $\mu_N$ represents the nucleon chemical potential. The vacuum contribution of the quark sector is given by
\begin{eqnarray}
\label{eq:NJL10}    
\mathscr{E}_V &=& 12i \int \frac{d^4k}{(2\pi)^4} \ln \Bigg[ \frac{k^2 -M_q^2 + i\epsilon}{k^2 -
M_0^2 + i \epsilon}\Bigg] \nonumber \\
&+& \frac{\left( M_q-m_q\right)^2}{4G_\pi} - \frac{\left(
M_0-m_q\right)^2}{4G_\pi},
\end{eqnarray}  
where $M_0$ denotes the vacuum constituent quark mass at zero baryon density.
Using the energy density obtained for SNM in the NJL model, the binding energy
per nucleon (the negative of the energy per nucleon relative to the free nucleon mass $M_{N0}$) can
be expressed as
\begin{eqnarray}
\frac{E}{A} &=& \frac{\mathscr{E}}{\rho_B} - M_{N0}.
\label{EBA}
\end{eqnarray}  
As discussed above, the stability of SNM within the NJL model
is ensured by reproducing the empirical energy per nucleon,
$E/A=-15.7~\mathrm{MeV}$, at
the saturation density $\rho_0=0.16~\mathrm{fm}^{-3}$. We then investigate the in-medium properties
of the $K^{*+}$ strange vector meson by employing the quark properties obtained in SNM from the NJL model as inputs.
\begin{table*}[t]
	\begin{ruledtabular}
		\renewcommand{\arraystretch}{2.0}
		\caption{In-medium modifications of the constituent quark mass (denoted by $M_q^*$),
vector meson masses,
		decay constants, meson-quark coupling constant, and magnetic moments, quadrupole
moment, charge radius, quark condensate, and nucleon effective mass calculated in the NJL model.
All units are in MeV.
The units of $Q_{K^{*+}}^*$ and $\big< r_{K^{*+}}^* \big>$ are in fm$^2$ and fm,
respectively.}
		\label{tab:NJL2}
		\begin{tabular}{ccccccccccccc}
		  $\rho_B/\rho_0$  & $M_q^*$ & $M_s$ & $m_{\rho^+}^*$ & $g_{\rho^{+}qq}$&$m_{K^{*+}}^*$ &$g_{\rm K^{*+} qq}^{*}$ & $\mu_{K^{*+}}^* (\mu_N)$ & $Q_{K^{*+}}^*$ &$\big< r_{K^{*+}}^*\big>$ &  $-\langle \bar{u}u \rangle^{*1/3}$ & $M_N^*$\\ 
\hline  
		 $0.00$ &  400  & 611 & 770 & 2.638 & 935 & 2.857  & 2.528  & $-0.053$ &  0.664 & 171 & 934.36 \\
		 $0.50$ &  360 & 611 & 726& 2.536 & 923&  2.776  & 2.641  & $-0.057$ & 0.703 & 165 & 842.54\\
		 $1.00$ &  328 & 611 & 693 & 2.459 & 917 & 2.708  & 2.783  &$-0.061$ & 0.735 & 160 & 778.03\\
		 $1.50$ &  306 & 611 & 673 & 2.409 & 914  &  2.660  & 2.856  & $-0.063$ & 0.757 & 156 & 738.98\\
          $2.00$ &  289 & 611 & 658 & 2.371 & 912  & 2.623   & 2.909 & $-0.065$ & 0.775 &  153 & 715.34\\
		\end{tabular}
		\renewcommand{\arraystretch}{1}
	\end{ruledtabular}
\end{table*}

%================================================================
\section{In-medium EMFFs}
\label{sec:ffv}
%================================================================
The in-medium electromagnetic current matrix element of the $K^{*+}$ meson
can be decomposed into three independent Lorentz structures, characterized by the charge, magnetic,
and quadrupole form factors,
\begin{eqnarray}
    \label{eq:vff1}
   \mathcal{J}_{K^{*+}}^{{*}\mu, \alpha \beta} (p'^{},p^{}) &=& \Big[ g^{\alpha \beta}
   F^{*}_{1K^{*+}} (Q^2) -\frac{q^{\alpha } q^{\beta }}{2 m_{K^{*+}}^{2}} F^{*}_{2K^{*+}} (Q^2) \Big]
\nonumber \\
   &\times& \left( p'^{} + p^{}\right)^\mu \nonumber \\
   &-& \Big[ q^{\alpha } g^{\mu \beta} - q^{\beta } g^{\mu \alpha}\Big] F^{*}_{3 K^{*+}} (Q^2),
\end{eqnarray}
where the Lorentz indices $\alpha$ and $\beta$ denote the polarization states of
the incoming and outgoing $K^{*+}$ meson, respectively, while $\mu$ corresponds to the photon index. The in-medium EMFFs $F^{*}_{1K^{*+}}(Q^2)$, $F^{*}_{2K^{*+}}(Q^2)$,
and $F^{*}_{3K^{*+}}(Q^2)$ characterize the internal electromagnetic structure of the $K^{*+}$ meson in the nuclear medium. In the Sachs representation, these form factors can be
expressed in terms of the in-medium charge (electric) form factor $G_C^{*K^{*+}}(Q^2)$, which
describes the charge distribution, the magnetic form factor $G_M^{*K^{*+}}(Q^2)$, which is related
to the magnetization distribution, and the quadrupole form factor $G_Q^{*K^{*+}}(Q^2)$, which
provides information on the spatial deformation of the $K^{*+}$ meson.
These form factors are defined as
\begin{eqnarray}
    \label{eqvff2}
    G^{*K^{*+}}_Q(Q^2) &=& F^{*}_{1K^{*+}} (Q^2) + A F^{*}_{2K^{*+}} (Q^2) \nonumber \\
    &- &F^{*}_{3K^*} (Q^2), \\
    G^{*K^{*+}}_C (Q^2) &=& F^*_{1K^{*+}} (Q^2) + \tfrac{2}{3} \eta\, G^{*K^{*+}}_Q (Q^2), \\
    G^{*K^{*+}}_M (Q^2) &=& F^*_{3K^{*+}} (Q^2),
\end{eqnarray}
where $A = (1 + \eta)$ with $\eta = Q^{2}/(4 m_{K^{*+}}^{2})$. It is worth noting
that all form factors are dimensionless.
\begin{figure}[b]
\centering
\includegraphics[width=1\columnwidth]{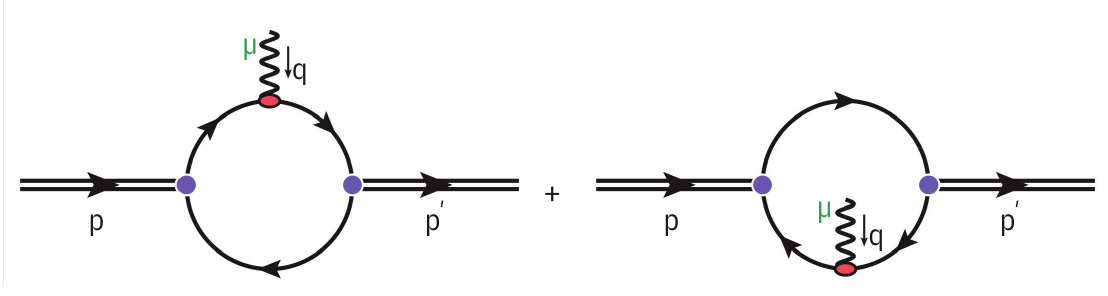} 
\caption{\label{fig6a} 
Two leading contributions to the electromagnetic current of the $K^{*+}$ meson.
The purple-shaded circles indicate the incoming and outgoing Bethe--Salpeter amplitudes, while the
red-shaded oval denotes the dressed quark--photon vertex.}
\end{figure}

Within the NJL framework, the electromagnetic current matrix element of the $K^{*+}$ meson
is obtained from the two leading diagrams depicted in Fig.~\ref{fig6a} and
takes the form
\begin{eqnarray}
    \label{eqvff2a}
    \mathcal{J}^{*\mu} (p',p) &=& i \int \frac{d^4k}{(2\pi)^4} \nonumber \\
    &\times& \mathrm{Tr} \Big[ \bar{\Gamma} S^{*}_{\ell}(p'+k) \Lambda^{*\mu}_{ \gamma Q} (p',p)
    \nonumber \\
    &\times& S^{*}_{\ell}(p +k) \Gamma S^{*T}_{s}(-k) \Big],
\end{eqnarray}
where $\mathrm{Tr}$ denotes the trace over Dirac, color, and isospin indices, and the superscript
$T$ represents the matrix transpose operation. The Bethe--Salpeter amplitudes of the $K^{*+}$ meson, corresponding to Eq.~(\ref{eqvff2a}) are given by
\begin{eqnarray}
    \Gamma_{K^{*+}}^\mu &=& g_{K^{*+}qq} \gamma^\mu \lambda_a,
\end{eqnarray}
where $\lambda_a=(\lambda_4+i\lambda_5)/2$ for the $K^{*+}$ meson, $\lambda_a=(\lambda_4-i\lambda_5)/2$ for the $K^{*-}$ meson, and $\lambda_a=(\lambda_6+i\lambda_7)/2$ for the $K^{*0}$ meson, with $\lambda_4$, $\lambda_5$, $\lambda_6$, and $\lambda_7$ denoting the Gell-Mann matrices that form part of the set $\lambda_i$ defined in Eq.~(\ref{eq:vacNJL1}). The quantity $\Lambda^{*\mu}_{\gamma Q}(p',p)$ in Eq.~(\ref{eqvff2a}) denotes the in-medium dressed quark--photon vertex, which satisfies the inhomogeneous Bethe--Salpeter equation,
\begin{eqnarray}
    \label{eqvff2b}
    \Lambda_i^{\big(\mathrm{BSE}\big) *\mu } (Q^2) &=& \gamma^\mu F^{*}_{1i} (Q^2)
    + \frac{i\sigma^{\mu \nu} q_\nu}{2M^{*}} F^{*}_{2i} (Q^2),~~~
\end{eqnarray}
where $i=(\rho,\omega)$ specifies the vector-isovector and vector-isoscalar channels, respectively.
The bare quark limit is recovered when the dressed quark form factors reduce to
$F^{*}_{1\omega}(Q^2)=1$ and $F^{*}_{2\rho}(Q^2)=F^{*}_{2\omega}(Q^2)=0$. Hence, the quark Pauli
form factor vanishes, $F^{*}_{2Q}(Q^2)=0$.

Finally, after deriving the electromagnetic current of the $K^{*+}$ meson within
the NJL model, as given in Eq.~(\ref{eqvff2a}), and matching it to the general current decomposition
in Eq.~(\ref{eq:vff1}), the $K^{*+}$ meson form factors can be extracted. By
consistently incorporating the full structure of the dressed quark propagators, the resulting
compact expressions for the $K^{*+}$ meson form factors can be written as
\begin{eqnarray}
F^{*}_{jK^{*+}} (Q^2) &=& \big[ F^{*}_{\mathrm{1U}} (Q^2) + F^{*}_{\mathrm{1S}} (Q^2)\big]
f_{jK^{*+}}^{*V} (Q^2) \nonumber \\
&+& \big[ F^{*}_{\mathrm{2U}} (Q^2) + F^{*}_{\mathrm{2S}} (Q^2)\big] f_{jK^{*+}}^{*T} (Q^2), ~~~
\end{eqnarray}
where $j=1,2,3$, and the vector body form factor contributions to the $K^{*+}$ meson form factors in the Schwinger proper-time regularization scheme are given by
\begin{eqnarray}
    \label{eq:bodyff}
    f_{1K^{*+}}^{*V} (Q^2) &=& - \frac{N_c g_{K^{*+} qq}^{*2}}{8 \pi^2}
    \int_{\tau_{\mathrm{UV}}}^{\tau_{\mathrm{IR}}} d\tau \int_0^1 dx \int_{-x}^{x} dy \,\, A_{1V}
\nonumber \\
    &\times& \exp\big[ -\tau (E_{1V}) \big] \nonumber \\
    &+& \frac{N_c g_{K^{*+} qq}^{*2}}{4\pi^2} \int_{\tau_{\mathrm{UV}}}^{\tau_{\mathrm{IR}}}
    \frac{d\tau}{\tau} \int_0^1 dx \nonumber \\
    &\times& \exp \big[ -\tau(E_{2V} )\big], 
\end{eqnarray}
\begin{eqnarray}
    f_{2K^{*+}}^{*V} (Q^2) &=& \frac{N_c g_{K^{*+} qq}^{*2} m_{K^{*+}}^{2}}{4\pi^2}
    \int_{\tau_{\mathrm{UV}}}^{\tau_{\mathrm{IR}}} d\tau \int_0^1 dx \int_{-x}^x dy \nonumber \\
    &\times& \big( x^2 -y^2 \big) \big( 1-x \big) \exp\Big[ -\tau \big( E_{1V} \big) \Big], 
\end{eqnarray}
\begin{eqnarray}
    f_{3K^{*+}}^{*V} (Q^2) &=& -\frac{N_c g_{K^{*+}qq}^{*2}}{8 \pi^2}
    \int^{\tau_{\mathrm{IR}}}_{\tau_{\mathrm{UV}}} d\tau \int_0^1 dx \int_{-x}^{x} dy \,\, A_{2V}
\nonumber \\
    &\times& \exp\Big[ -\tau \big( E_1\big) \Big] \nonumber \\
    &+& \frac{3 N_c g_{K^{*+}qq}^{*2}}{4 \pi^2}  \int_0^1 dx
    \int_{\tau_{\mathrm{UV}}}^{\tau_{\mathrm{IR}}} \frac{d\tau}{\tau} \nonumber \\
    &\times& \exp\Big[ -\tau \big( E_{2V} \big)\Big],
\end{eqnarray}
where the variables $A_1$, $A_2$, $E_1$, and $E_2$ are defined as follows
\begin{eqnarray}
    A_{1V} &= & \Big[ \frac{2}{\tau} \big(1-x\big) - x m_{K^{*+}}^{*2} + (M^{*}_{s} -M^{*}_{l})
    \bigl\{2M^{*}_{s} \nonumber \\
    &-& x (M^{*}_{s} - M^{*}_{l})\bigr\} \Big],  \\
    A_{2V} &=& \Big[ \frac{2}{\tau} (x+1)-x(1+2x) m_{K^*}^{*2} - (x^2 -y^2) \frac{Q^2}{2}  \nonumber\\
    &+& (M^{*}_{s} - M^{*}_{l}) \bigl\{x (M^{*}_{s} + M^{*}_{l}) + 2 M^{*}_{s} \bigr\}\Big],  \\
    E_{1V} &=& \Big[ (x^2-x) m_{K^{*+}}^{*2} + \frac{1}{4} (x^2 -y^2) Q^2 + xM_{l}^{*2} \nonumber \\
    &+& (1-x) M_{s}^{*2}\Big],  \\
    E_{2V} &=& (x-x^2)Q^2 + M_{l}^{*2},
\end{eqnarray}
 while the in-medium modifications of the tensor body form factors read 
\begin{eqnarray}
    \label{eq:bodyffb}
    f_{1K^*}^{*T} (Q^2) &=& - \frac{N_c g_{K^{*+} qq}^{*2} Q^2}{64 \pi^2 M^{*}_{q_1}}
    \int_0^1 dx \int_{-x}^{x} dy \int_{\tau_{\mathrm{UV}}}^{\tau_{\mathrm{IR}}} d\tau A_{1T}
\nonumber \\
    &\times& \exp \Big[ -\tau \big( E_{1V} \big)\Big],
\end{eqnarray}
\begin{eqnarray}
    f_{2K^*}^{*T} (Q^2) &=& \frac{N_c g_{K^{*+} qq}^{*2}m_{K^{*+}}^{*2}}{16\pi^2 M^{*}_{l}}
    \int_0^1 d
    x \int_{-x}^{x} dy \int_{\tau_{\mathrm{UV}}}^{\tau_{\mathrm{IR}}} d\tau A_{2T} \nonumber \\
    &\times& \exp \Big[ -\tau \big( E_{1V} \big) \Big], 
\end{eqnarray}
\begin{eqnarray}
    f_{3 K^*}^{*T} (Q^2) &=& - \frac{N_c g_{K^{*+}qq}^{*2}}{64 \pi^2 M^{*}_{l}} \int_0^1 dx \int_{-x}^{x} dy \int_{\tau_{\mathrm{UV}}}^{\tau_{\mathrm{IR}}} d\tau A_{3T} \nonumber \\
    &\times& \exp \Big[ -\tau \big( E_{1V}\big)\Big] \nonumber \\
    &+& \frac{N_c g_{K^{*+} qq}^{*2}}{16\pi^2M^{*}_{l}} \int_0^1 dx \int_{\tau_{\mathrm{UV}}}^{\tau_{\mathrm{IR}}} \frac{d\tau}{\tau} A_{4T} \nonumber \\
    &\times& \exp \Big[ -\tau \big( E_{2V} \big) \Big], 
\end{eqnarray}
where the variables $A_{1T}$, $A_{2T}$, $A_{3T}$, and $A_{4T}$ are defined, respectively, by
\begin{eqnarray}
    A_{1T} &=& \bigl[(1-x)M^*_{s} + x M^*_{l}\bigr], \\
    A_{2T} &=& \bigl[ M^*_{s} -xM^*_{l} + (x^2 -2x)(M^*_{s} - M^*_{l})\bigr], \\
    A_{3T} &=& \bigl[ xM^*_{l} Q^2 - 2M^*_{s} m_{K^*}^{*2} + 2(M^*_{s} - M^*_{l})\bigr], ~~\\
    A_{4T} &=& \bigl[M^*_{l} -(M^*_{s} -M^*_{l})\bigr].
\end{eqnarray}
It is important to note that, in the strange-quark sector, the tensor contribution vanishes when the virtual photon couples directly to the strange quark, as $F^{*}_{2S}(Q^2)=0$. Namely, pion-cloud contributions are not included in the present calculation. The strange-quark tensor form factor $F^{*}_{2S}(Q^2)$ arises only when
pion-cloud corrections are incorporated through the inhomogeneous Bethe--Salpeter equation.
The dressed quark form factors obtained from the inhomogeneous BSE can be defined as
\begin{eqnarray}
\label{eq:vff4}
F^*_{1U} (Q^2) &=& \tfrac{1}{6} F^*_{1\omega} (Q^2) + \tfrac{1}{2} F^*_{1 \rho} (Q^2),  \\
F^*_{1D} (Q^2) &=& \tfrac{1}{6} F^*_{1\omega} (Q^2) - \tfrac{1}{2} F^*_{1 \rho} (Q^2), \\
F^*_{2U} (Q^2) &=& \tfrac{1}{6} F^*_{2\omega} (Q^2) + \tfrac{1}{2} F^*_{2 \rho} (Q^2),  \\
F^*_{2D} (Q^2) &=& \tfrac{1}{6} F^*_{2\omega} (Q^2) - \tfrac{1}{2} F^*_{2 \rho} (Q^2), \\
F^*_{1S} (Q^2) &=& e_s F^*_{1\phi} (Q^2).
\end{eqnarray}
The vanishing of the Pauli form factors, $F^{*}_{2\rho}(Q^2)=F^{*}_{2\omega}(Q^2)=0$, is a consequence of the interaction kernel employed in the present NJL truncation, which does not contain a tensor-tensor four-fermion channel. Hence, an anomalous magnetic-moment component is not dynamically generated by the corresponding inhomogeneous Bethe–Salpeter equation. This should be regarded as a truncation effect rather than a general QCD prediction. In more elaborate theoretical approaches, such as light-front constituent-quark models and Dyson–Schwinger/Bethe–Salpeter frameworks, relativistic spin-dependent interactions and momentum-dependent quark dressing can generate nonzero Pauli-type contributions. Their effect is expected to be less pronounced for the charge form factor, which is dominated by the Dirac component, but potentially more relevant for magnetic and, in particular, quadrupole observables because ($F_2(Q^2)$) enters explicitly into the spin-1 quadrupole form factor. Therefore, while the omission of ($F_2 (Q^2)$) is unlikely to qualitatively alter the dominant charge distribution obtained here, a quantitative assessment of its impact on magnetic and quadrupole moments requires a calculation with an extended interaction kernel. 

The static electromagnetic properties of the $K^{*+}$ meson can be extracted from the corresponding
Sachs form factors. In particular, the magnetic moment $\mu^{*}_{K^{*+}}$, quadrupole
moment $\mathcal{Q}^{*}_{K^{*+}}$, and mean-square charge radius $\langle r_{K^{*+}}^{*2}\rangle$
in the nuclear medium are obtained by taking the limit $Q^2\rightarrow0$ as
\begin{eqnarray}
    \mu^{*}_{K^{*+}} &=& G^{*{K^{*+}}}_M (Q^2 =0)\,\, \frac{M^*_N}{m^*_{K^{*+}}}, \\
    \mathcal{Q}^*_{K^{*+}} &=& \frac{G^*_{Q} (Q^2=0)}{m^{ *2}_{K^{*}}}, \\
    \big< r_{K^*+}^{*2} \big> &=& -6 \frac{\partial G^*_C (Q^2)}{\partial Q^2} \Bigg|_{Q^2 =0},
\end{eqnarray}
where $\mu^*_{K^{*+}}$ denotes the in-medium magnetic moment of the $K^{*+}$ meson, expressed
in units of the nuclear magneton $\mu_N$ (free space), with $M_N^*$ denoting the in-medium
nucleon mass. The charge conservation constraint requires the normalization condition
$G_C^{*K^{*+}}(0)=1$. For comparison, we also present the corresponding results for the $\rho^+$
meson.

%================================================================
\section{Numerical Result and Discussion} \label{sec:MR}
\begin{figure}[t]
	\centering
	\includegraphics[width=1\columnwidth]{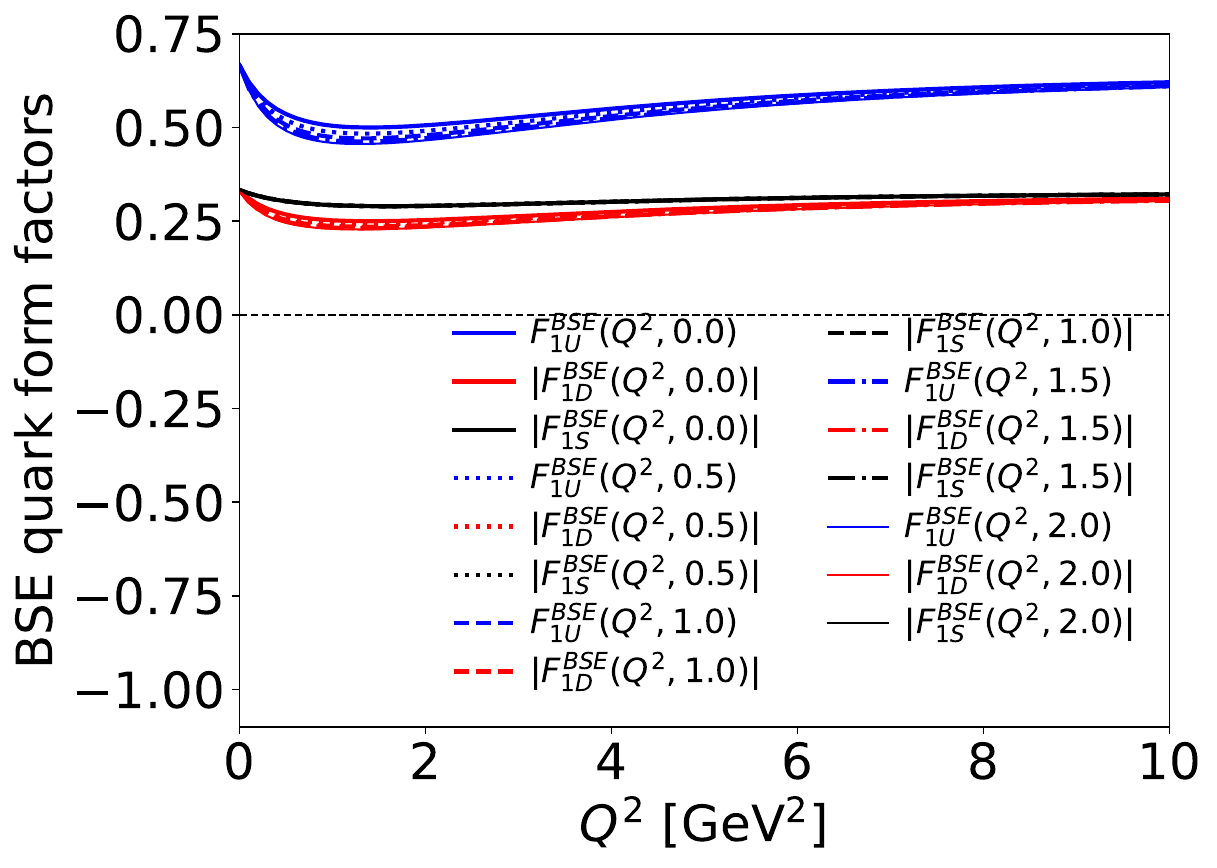}
 	\caption{\label{fig1} 
Results for the BSE dressed form factors of the up, down, and strange quarks at different
nuclear densities, $\rho/\rho_0$, as functions of the squared four-momentum
transfer, $Q^2$. For clarity, the down- and strange-quark BSE form factors are plotted as
$-F_{1D}^{\rm BSE}(Q^2)$ and
$-F_{1S}^{\rm BSE}(Q^2)$, respectively, so that their displayed form factors are positive.}
\end{figure}

\begin{figure*}[t]
	\centering
	\includegraphics[width=0.6\textwidth]{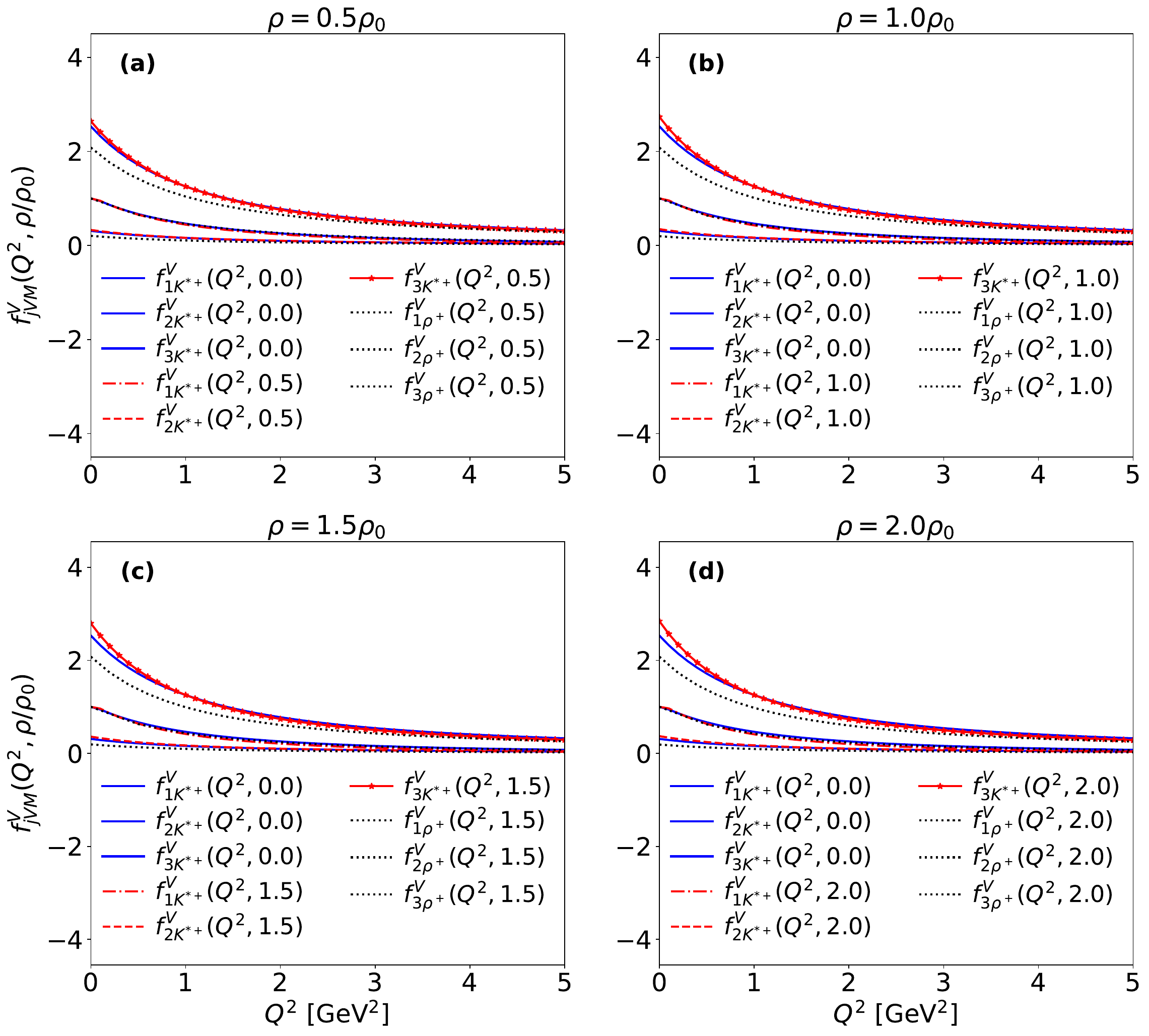}
 	\caption{\label{fig2} Vector-meson form factors $f_{1K^{*+}}(Q^2)$, $f_{2K^{*+}}(Q^2)$, and
$f_{3K^{*+}}(Q^2)$ as functions of the squared four-momentum transfer $Q^2$ at different
nuclear densities $\rho/\rho_0$.}
\end{figure*}

\begin{figure*}[!]
	\centering
	\includegraphics[width=0.6\textwidth]{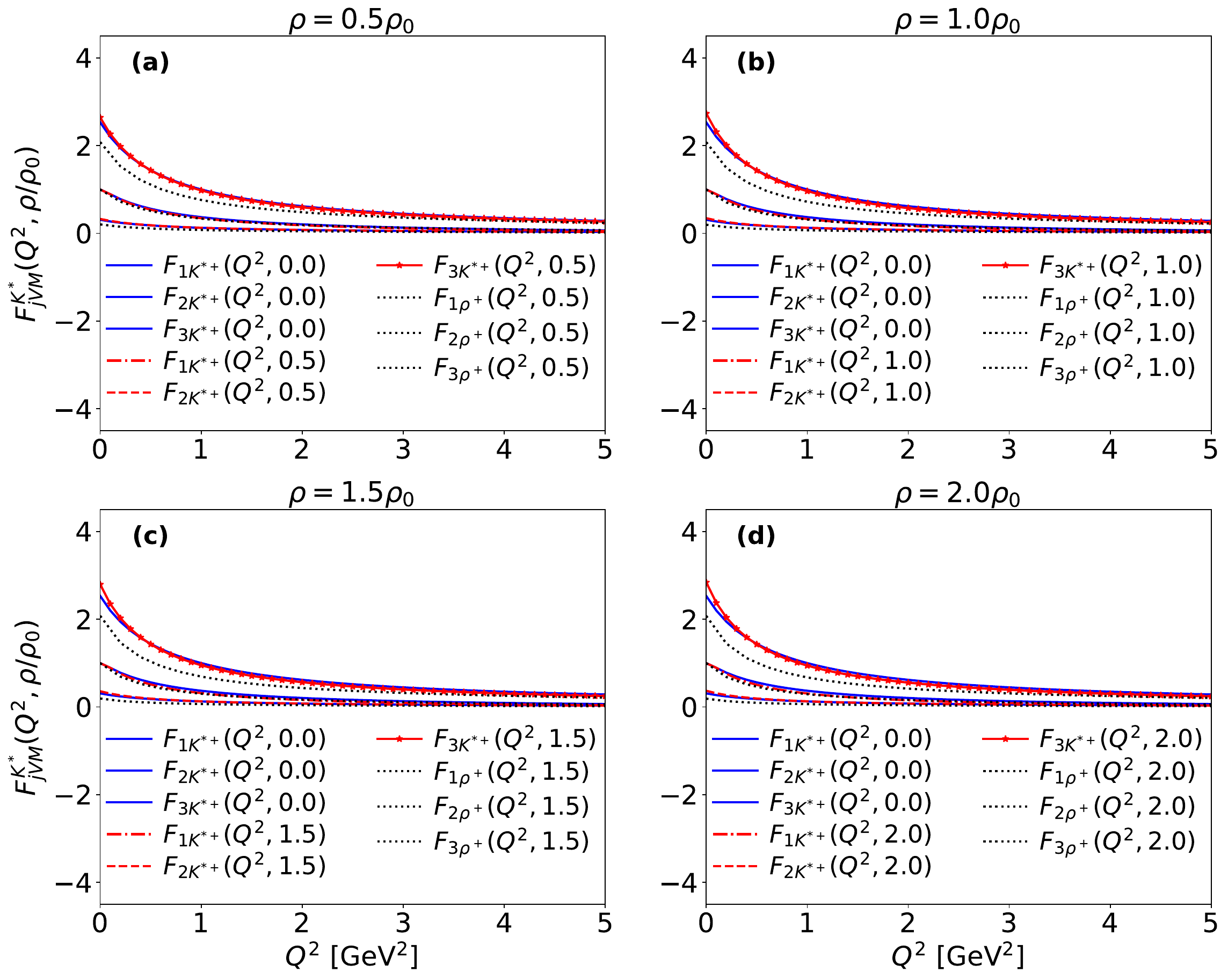}
 	\caption{\label{fig3} Results for the BSE-dressed form factors $F_{1K^{*+}}(Q^2)$,
 	$F_{2K^{*+}}(Q^2)$, and $F_{3K^{*+}}(Q^2)$ for different nuclear densities $\rho/\rho_0$ as  a
function of the squared four-momentum transfer $Q^2$.}
\end{figure*}

\begin{figure*}[t]
	\centering
	\includegraphics[width=0.6\textwidth]{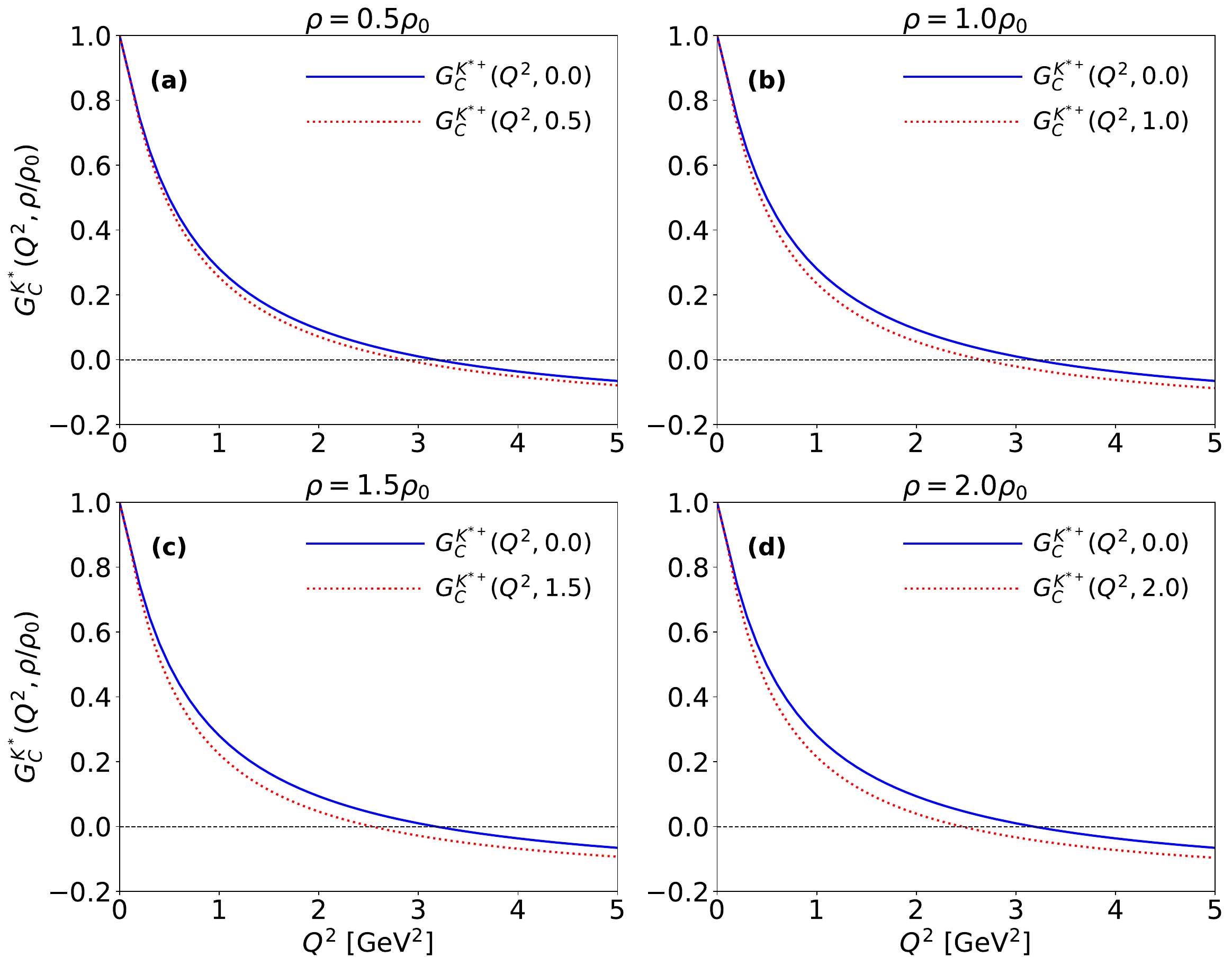}
 	\caption{\label{fig4} Results for the charge (electric) form factors for different
 	nuclear densities $\rho/\rho_0$ as a function of the squared four-momentum transfer
$Q^2$.}
\end{figure*}

\begin{figure*}[!]
	\centering
	\includegraphics[width=0.6\textwidth]{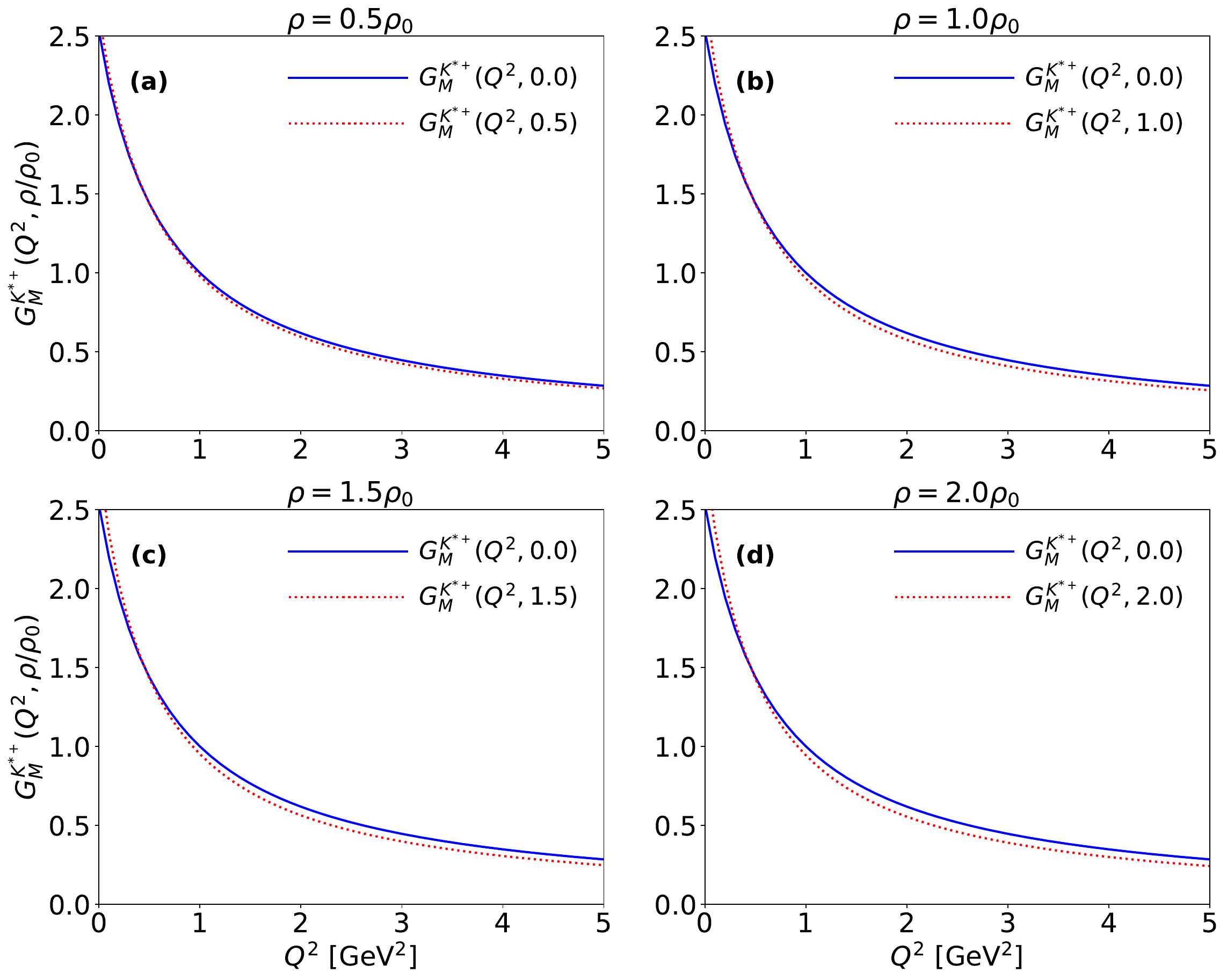}
 	\caption{\label{fig5}
Results for the magnetic form factors for different nuclear densities $\rho/\rho_0$ as a function
of the squared four-momentum transfer $Q^2$.}
\end{figure*}

\begin{figure*}[!]
	\centering
	\includegraphics[width=0.6\textwidth]{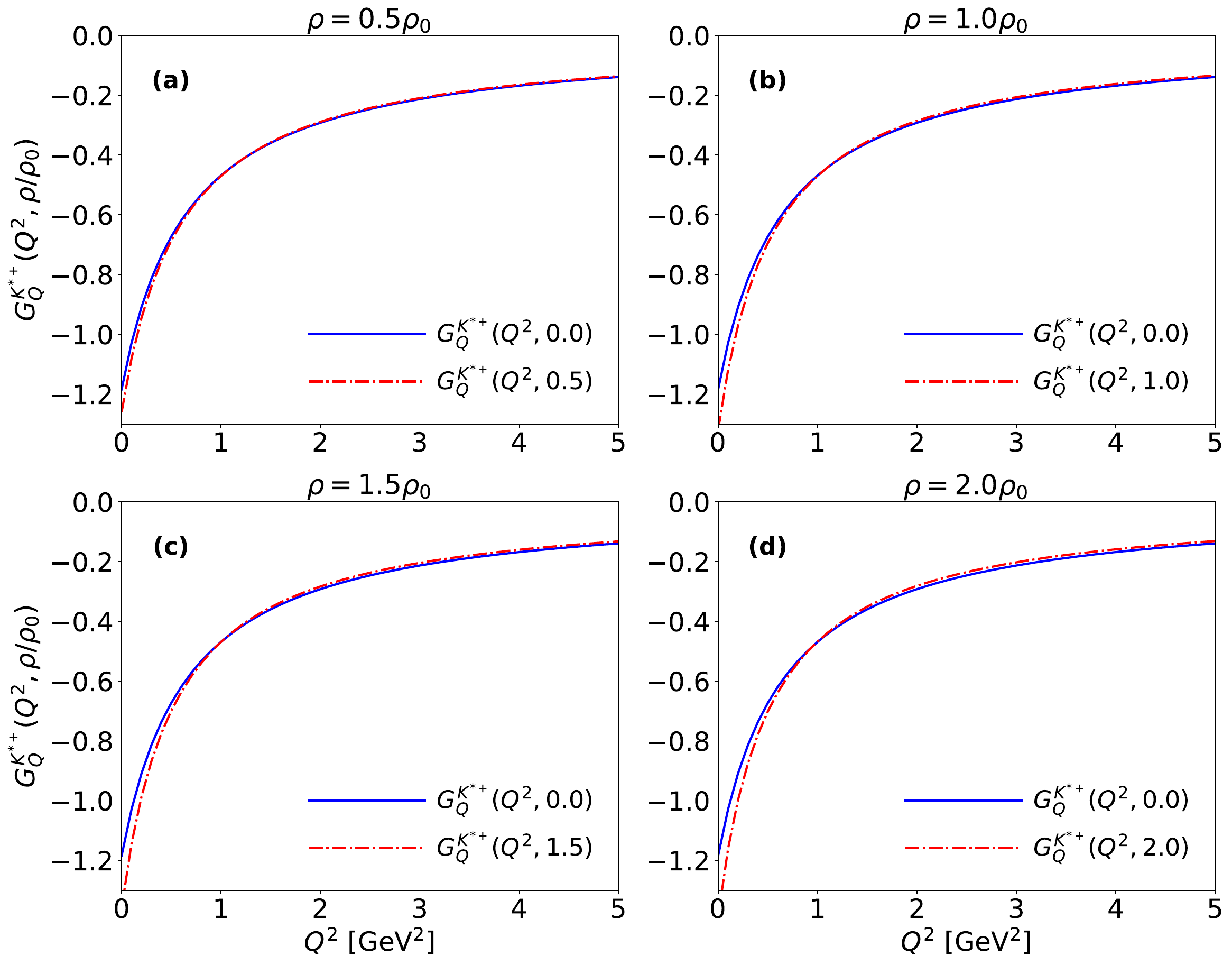}
 	\caption{\label{fig6} Results for the quadrupole form factors for different nuclear densities
 	$\rho/\rho_0$ as a function of the squared four-momentum transfer $Q^2$.}
\end{figure*}

%================================================================
The calculated in-medium EMFFs of the $K^{*+}$ meson, together with its charge radius $r_{K^{*+}}$,
are shown in Figs.~\ref{fig1}--\ref{fig6}. The density dependence of the
charge, magnetic, and quadrupole form factors, $G_C^*(Q^2)$, $G_M^*(Q^2)$, and $G_Q^*(Q^2)$,
respectively, is also presented.

The NJL model parameters employed in this work are the coupling constants
$G_\pi$, $G_\omega$, $G_a$, $G_s$, and $G_\rho$, following
Refs.~\cite{Gifari:2024ssz,Hutauruk:2021kej,Hutauruk:2016sug,Hutauruk:2018zfk}. Within the Schwinger
proper-time regularization scheme, the infrared cutoff is fixed at
$\Lambda_{\mathrm{IR}}=240~\mathrm{MeV}$, which is close to the QCD scale $\Lambda_{\mathrm{QCD}}$.
The dynamical light-quark mass is chosen as $M_{\ell} =400~\mathrm{MeV}$. The remaining model
parameters are determined by fitting to the physical observables $m_\pi=140~\mathrm{MeV}$,
$m_K=495~\mathrm{MeV}$, $M_N=934.36~\mathrm{MeV}$, $m_\rho=770~\mathrm{MeV}$, and the pion decay
constant $f_\pi=93~\mathrm{MeV}$. The resulting parameter set gives an ultraviolet cutoff
$\Lambda_{\mathrm{UV}}=645~\mathrm{MeV}$, pion-channel coupling constant
$G_\pi=19.04\times10^{-6}~\mathrm{MeV}^{-2}$, vector coupling constant
$G_\rho=11.04\times10^{-6}~\mathrm{MeV}^{-2}$, strange constituent quark mass
$M_s=611~\mathrm{MeV}$, and $m_{K^{*+}}=935~\mathrm{MeV}$. The current up (down)- and
strange-quark masses
are taken as $m_{u,d}=16~\mathrm{MeV}$ and $m_s=356~\mathrm{MeV}$, respectively, consistent with Refs.~\cite{Gifari:2024ssz,Hutauruk:2021kej,Hutauruk:2016sug,Hutauruk:2018zfk}.
The $\phi$ vector meson mass is taken as $m_\phi =$ 1001 MeV, the scalar diquark mass $M_{sd}
=$ 687 MeV, and the axial diquark mass $M_{ad} =$ 1027 MeV. Note that the coupling constants
$G_{s}$ and $G_a$ are determined by solving the Faddeev equation to reproduce the free-space nucleon
mass $M_N$ and axial coupling constant $g_A =$ 1.267, as used in
Ref.~\cite{Bentz:2001vc,Hutauruk:2025bjd}. 

The parameters associated with SNM are determined by fitting the negative of the binding energy per
nucleon $E/A=-15.7$ MeV at the normal nuclear density $\rho_0 =$
0.16 fm$^{-3}$ [see Eq.~(\ref{EBA})]. This results in $G_\omega = 6.03 \times 10^{-6}$ MeV$^{-2}$.
Using these parameters,
we compute the $K^{*+}$ meson--quark coupling using the expression given in
Eq.~(\ref{eq:vacuumNJL9}), we obtain $g_{K^{*+}qq}=2.857$ in free space, while the complete results
for the in-medium modifications of the coupling constants are presented in Table~\ref{tab:NJL2}.

We first present the in-medium modifications of the Bethe--Salpeter equation (BSE)
dressed form factors of the up, down, and strange valence quarks at different nuclear densities, as
shown in Fig.~\ref{fig1}. For the up quark, the BSE dressed form factor satisfies the normalization
condition $F_{1U}^{*\mathrm{BSE}}(0)=e_u={2}/{3}$, both in free space and in nuclear matter, as
required by electric charge conservation. In the asymptotic limit, $Q^2\rightarrow\infty$, the form
factor approaches $F_{1U}^{*\mathrm{BSE}}(Q^2)\rightarrow e_u={2}/{3}$, consistent with the
expected asymptotic behavior~\cite{Cloet:2014rja}. Physically, this reflects the fact that the
dressing effects become negligible at large momentum transfer, and the up quark behaves as a
pointlike particle. In the intermediate momentum-transfer region, $0<Q^2<8~\mathrm{GeV}^2$, the BSE
dressed form factor is suppressed with increasing nuclear density, indicating significant medium
modifications of the quark dressed form factors.

A similar density dependence is observed for the down and strange quarks, as shown
in Fig.~\ref{fig1}. The only differences arise from their electric charges, which determine the
normalization and asymptotic limits of the corresponding form factors:
$F_{1D}^{*\mathrm{BSE}}(0) = F_{1D}^{*\mathrm{BSE}}(\infty) = e_d = - 1/3$,
and
$F_{1S}^{*\mathrm{BSE}}(0)= F_{1S}^{*\mathrm{BSE}}(\infty) = e_s = -1/3$, respectively.

In Fig.~\ref{fig2}(a)--(d), we present the body form factors of the $K^{*+}$ meson
at different nuclear densities. For comparison, the corresponding body form factors
of the $\rho^+$ meson are also shown. As illustrated in Fig.~\ref{fig2}(a), the body form factor of
the $K^{*+}$ meson is suppressed with increasing nuclear density. Compared with the
corresponding $\rho^+$ meson result, $f_{3K^{*+}}^{V}(Q^2,0.5)$ exhibits a more
moderate decrease with increasing the squared four-momentum transfer $Q^2$ than
$f_{3\rho^{+}}^{V}(Q^2,0.5)$,
whereas $f_{1K^{*+}}^{V}(Q^2,0.5)$ displays a weaker $Q^2$ dependence than
$f_{1\rho^{+}}^{V}(Q^2,0.5)$. At $Q^2=0$, the normalization condition $f_{1K^{*+}}^{V}(0,0.5)=1$ is
satisfied, ensuring charge conservation, as in free space. By contrast, the in-medium modifications
of the body form factor $f_{2K^{*+}}^{V}(Q^2,0.5)$ are considered relatively small over the entire
momentum-transfer region studied. Nevertheless, it differs noticeably from the
corresponding body form factor of the $\rho^+$ meson, particularly for the squared momentum transfer
$Q^2\lesssim 3~\mathrm{GeV}^2$.

A similar trend is observed for the in-medium body form factors $f_{1K^{*+}}^{V}(Q^2)$,
$f_{2K^{*+}}^{V}(Q^2)$, and $f_{3K^{*+}}^{V}(Q^2)$ at nuclear densities $\rho/\rho_0=1.0$, 1.5, and
2.0, as shown in Figs.~\ref{fig2}(b), (c) and (d), respectively. Overall,
the body form factor
$f_{3K^{*+}}^{V}(Q^2)$ is suppressed with increasing nuclear density, with the suppression becoming
more pronounced at higher densities. In contrast, the in-medium body form factor
$f_{2K^{*+}}^{V}(Q^2)$ is enhanced relative to its free-space value, and the enhancement increases
with nuclear density increasing, particularly in the low-$Q^2$ region.

In addition to the body form factors, we present the BSE-dressed quark form factors
$F_{1K^{*+}}(Q^2)$, $F_{2K^{*+}}(Q^2)$, and $F_{3K^{*+}}(Q^2)$ at different nuclear densities in
Fig.~\ref{fig3}(a)--(d). As the nuclear density increases, the form factors $F_{1K^{*+}}(Q^2)$ and
$F_{3K^{*+}}(Q^2)$ exhibit a progressively stronger suppression, as shown in
Figs.~\ref{fig3}(a)--(d). In contrast, the form factor $F_{2K^{*+}}(Q^2)$ is enhanced in the nuclear
medium relative to its free-space value, with the enhancement becoming more pronounced at higher
nuclear densities. Compared with the corresponding BSE-dressed form factors of the $\rho^{+}$ vector
meson, those of the $K^{*+}$ strange vector meson are slightly larger over the momentum-transfer
region studied. This behavior can be attributed to the presence of the strange quark, whose
larger dynamical
mass modifies the internal quark dynamics and consequently the electromagnetic structure of the
$K^{*+}$ meson.

The charge (electric) form factor of the $K^{*+}$ meson at a nuclear density of
$\rho/\rho_0=0.5$ is presented in Fig.~\ref{fig4}(a). Relative to its free-space counterpart,
$G_C^{K^{*+}}(Q^2,0.5)$ exhibits a softer $Q^2$ dependence. The normalization condition,
$G_C^{K^{*+}}(0)=1$, is satisfied both in free space and at $\rho/\rho_0=0.5$, as required by charge
conservation. In addition, the in-medium charge form factor exhibits a zero crossing at
approximately $Q^2\simeq3.0~\mathrm{GeV}^2$.

Figure~\ref{fig4}(b) presents the corresponding results at $\rho/\rho_0=1.0$. Compared with the
free-space result and that at $\rho/\rho_0=0.5$, the charge form factor is further suppressed
throughout the momentum-transfer range studied. The normalization condition
$G_C^{K^{*+}}(0)=1$ remains satisfied, while the zero-crossing point shifts slightly toward lower
$Q^2$.

The same trend persists at higher nuclear densities. As shown in Fig.~\ref{fig4}(c), the charge
form factor at $\rho/\rho_0=1.5$ exhibits a softer $Q^2$ dependence than those at $\rho/\rho_0=0.5$
and 1.0. At $\rho/\rho_0=2.0$, shown in Fig.~\ref{fig4}(d), the suppression becomes even more
pronounced. Similar to the lower-density cases, the zero-crossing point continues to move toward
lower values of $Q^2$ as the nuclear density increases.

Overall, the charge form factor $G_C^{K^{*+}}(Q^2)$ is gradually suppressed with increasing nuclear
density, and the suppression becomes more significant at higher nuclear densities. In free
space, the charge form factor exhibits a zero crossing at approximately
$Q^2\simeq3.0~\mathrm{GeV}^2$. As the nuclear density increases, the position of the zero crossing
shifts systematically toward lower momentum transfer $Q^2$, reflecting the medium
modifications of the charge (electric) form factors of the $K^{*+}$ meson.

The results for the magnetic form factor of the $K^{*+}$ meson at nuclear densities
$\rho/\rho_0=$ 0.0, 0.5, 1.0, 1.5, and 2.0 are presented in Figs.~\ref{fig5}(a)--(d). Similar to
the behavior of the charge form factor, $G_C^{K^{*+}}(Q^2)$, the magnetic form factor
$G_M^{K^{*+}}(Q^2)$ is gradually suppressed with increasing nuclear density. The in-medium
modifications become more pronounced at higher nuclear densities, as illustrated in
Fig.~\ref{fig5}(d). Note that our free-space results for
$G_M^{K^{*+}}(Q^2)$ are in good agreement with the corresponding results reported in
Ref.~\cite{Hawes:1998bz,Bhagwat:2006pu}.

Figures~\ref{fig6}(a)--(d) present the quadrupole form factor, $G_Q^{K^{*+}}(Q^2)$, of the $K^{*+}$
meson at different nuclear densities. We find that the quadrupole
form factor $G_Q^{K^{*+}}(0)$ is negative for all nuclear densities considered, consistent with the
free-space result reported in Ref.~\cite{Hawes:1998bz,Bhagwat:2006pu}. Furthermore, the
medium-induced modifications of $G_Q^{K^{*+}}(Q^2)$ are most pronounced in the low-momentum-transfer
region, $Q^2 \lesssim 1~\mathrm{GeV}^2$, where the deviation from the corresponding free-space
result is largest. At higher momentum transfer, the in-medium effects gradually diminish, and the
form factor approaches its free-space value.

\begin{figure}[t]
	\centering
	\includegraphics[width=0.8\columnwidth]{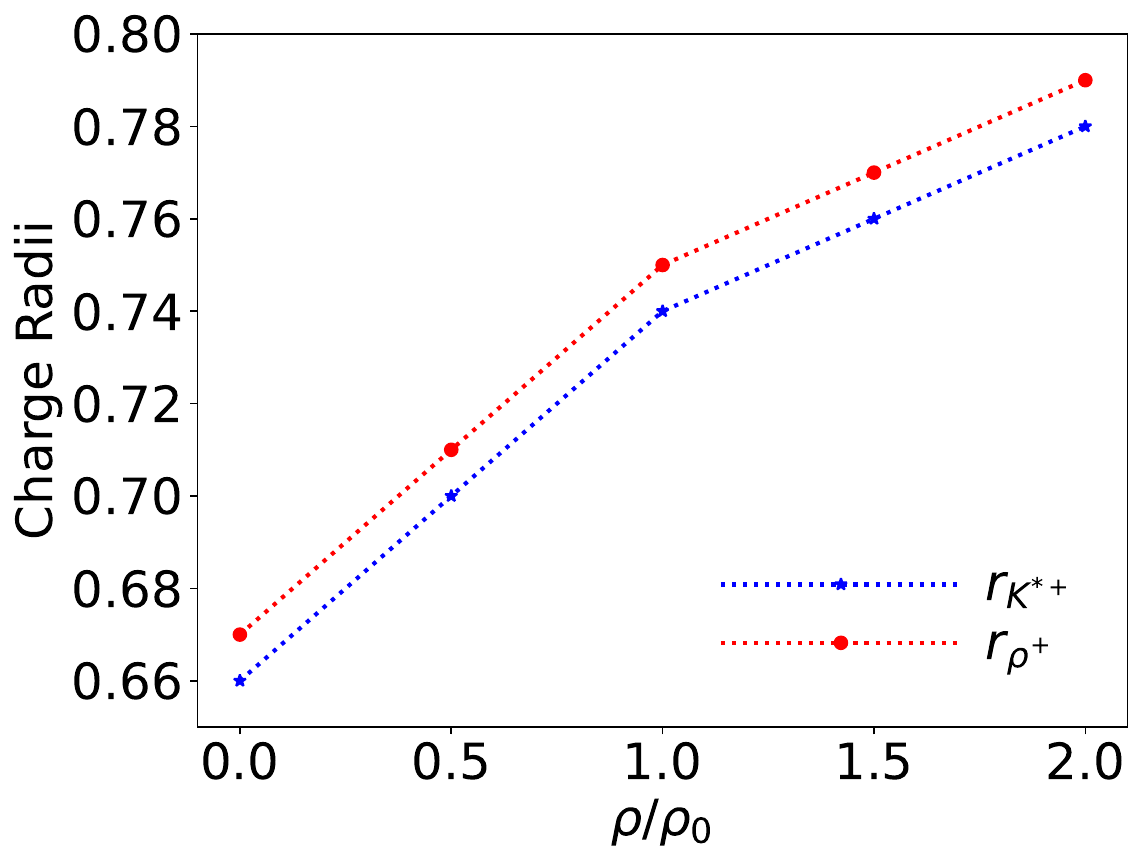}
 	\caption{\label{fig7} Charge radius of the $K^{*+}$ meson as a function of nuclear density, compared with that of the $\rho^+$ meson.}
\end{figure}

Figure~\ref{fig7} presents the charge radius of the $K^{*+}$ meson as a function of
the nuclear density, together with the corresponding result for the $\rho^{+}$ vector meson for
comparison. We find that the charge radius of the $K^{*+}$ meson increases monotonically with
increasing nuclear density, exhibiting a trend similar to that of the $\rho^{+}$ meson.
Quantitatively, however, the two mesons differ: the charge radius of the $K^{*+}$ meson is found to
be smaller than that of the $\rho^{+}$ meson over the entire nuclear density range. This is consistent with the intuitive expectation that heavier mesons are more compact and, hence, more tightly bound.
The complete values of the charge radius of the $K^{*+}$ meson for different nuclear densities are
summarized in Table~\ref{tab:NJL2}.

%================================================================
\section{Summary and Conclusion} \label{sec:summary}
%================================================================
In summary, we have studied the electromagnetic structure of the $K^{*+}$ meson
within the covariant NJL model with Schwinger proper-time regularization. The regularization scheme
provides ultraviolet regularization, while eliminating unphysical quark production at thresholds
associated with the absence of confinement in the NJL model, which is given by
the infrared cutoff limit. We have calculated the charge, magnetic, and quadrupole form factors,
$G_C^{*K^{*+}}(Q^2)$, $G_M^{*K^{*+}}(Q^2)$, and $G_Q^{*K^{*+}}(Q^2)$, together with the corresponding
charge radii of the $K^{*+}$ mesons in SNM.

We have found that the in-medium modifications of the charge form factor, $G_C^{*K^{*+}}(Q^2)$,
are gradually suppressed with increasing nuclear density. In free space, $G_C^{*K^{*+}}(Q^2)$
exhibits a zero crossing at approximately $Q^2 \simeq 3.0~\mathrm{GeV}^2$. As the nuclear density
increases, the position of the zero crossing shifts toward lower momentum transfer, indicating a
significant
medium modification of the charge distribution of the $K^{*+}$ strange vector meson.

A similar trend is observed for the magnetic form factor, $G_M^{*K^{*+}}(Q^2)$, which is suppressed
with increasing nuclear density. The suppression becomes more pronounced at higher densities.
Furthermore, our free-space results for $G_M^{K^{*+}}(Q^2)$ are in good agreement with those
reported in Refs.~\cite{Hawes:1998bz,Bhagwat:2006pu}.

For the quadrupole form factor, $G_Q^{K^{*+}}(Q^2)$, we found that the quadrupole form factor,
$G_Q^{K^{*+}}(0)$, is negative for all nuclear densities considered,
consistent with the corresponding free-space result reported in
Refs.~\cite{Hawes:1998bz,Bhagwat:2006pu}. The in-medium modifications of $G_Q^{*K^{*+}}(Q^2)$ is
most pronounced in the low-momentum-transfer region, $Q^2 \lesssim 1~\mathrm{GeV}^2$.

Finally, we found that the charge radius of the $K^{*+}$ meson increases monotonically
with increasing nuclear density. Moreover, the charge radius of the $K^{*+}$ meson is
found to be smaller than that of the corresponding $\rho^{+}$ meson over the entire density
range and the squared four-momentum transfer range studied.
At normal nuclear density, we obtained
$r^{*}_{K^{*+}} = 0.74$ fm with $\mu^{*}_{K^{*+}} =$
2.78 $\mu_N$ and $\mathcal{Q}^{*}_{K^{*+}} = -0.06$ fm, as shown in Table~\ref{tab:NJL2}.

Future high-precision measurements and analyses at facilities such as BABAR,
PANDA, Belle II, Jefferson Lab, and the Electron--Ion Collider (EIC) will provide stringent tests of
theoretical predictions for the EMFFs of vector mesons, including the charge, magnetic, and
quadrupole form factors, over a broad range of momentum transfers. In particular, future measurements
involving nuclear targets will provide valuable insight into the in-medium modifications of the
electromagnetic structure of vector mesons.

%================================================================
\section*{Acknowledgements}
%================================================================
This work was supported by the World Premier International Research Center Initiative (WPI)
of Hiroshima University, MEXT, Japan. T.M. was supported in part by the PUTI Q1 Grant from
University of Indonesia under contract PKS-206/UN2.RST/HKP.05.00/2025. P.T.P.H. and K.T. were
partially supported by the RCNP Collaboration Research Network Program under Project No.~COREnet
057. K.T. was supported by the National Council for Scientific and Technological
Development – CNPq, Brazil, Processes No.~304199/2022-2 and No.~306866/2026-9, by the S\~{a}o Paulo
Research Foundation (FAPESP), Process No.~2023/07313-6 and No.~2026/01656-7, and by the Instituto
Nacional de Ci\^{e}ncia e Tecnologia - Nuclear Physics and Applications (INCT-FNA), Brazil, Process
No.~408419/2024-5.

%----------------------------------------------------------------
\bibliographystyle{elsarticle-num}
\bibliography{main}
%----------------------------------------------------------------
\end{document}